\documentclass[osajnl,preprint,showpacs,superscriptaddress,12pt]{revtex4-1} 
\usepackage{amsmath,amssymb,graphicx}

\usepackage{lmodern, geometry, epsfig, epsf, ulem}
\usepackage[T1]{fontenc}
\usepackage{longtable}

\newcommand{\ti}{\theta_i}

\newcommand{\x}{$X_1$} 
\newcommand{\y}{$Y_1$} 
\newcommand{\z}{$Z_1$}

\begin{document}

\title{Membrane based Deformable Mirror: Intrinsic aberrations and alignment issues}

\author{A. Raja Bayanna}\email{bayanna@prl.res.in}
\affiliation{Physical Research Laboratory, Udaipur Solar Observatory, Udaipur-313034, India.}
\author{Rohan E. Louis}
\affiliation{Leibniz-Institut f\"{u}r Astrophysik Potsdam (AIP), An der Sternwarte 16, 14482, Potsdam, Germany}
\author{S. Chatterjee}
\affiliation{Indian Institute of Astrophysics, Koramangala, Bangalore-560017, India and \\BGVS, Indian Institute of Science Campus, Bangalore, 560012, India}
\author{Shibu K. Mathew}
\author{P. Venkatakrishnan}
\affiliation{Physical Research Laboratory, Udaipur Solar Observatory, Udaipur-313034, India.}

\begin{abstract}A Deformable Mirror (DM) is an important component of an Adaptive Optics system.  It is known that an on-axis spherical/parabolic optical component, placed at an angle to the incident beam introduces defocus as well as astigmatism in the image plane. Although the former can be compensated by changing the focal plane position, the latter cannot be removed by mere optical re-alignment. Since the DM is to be used to compensate a turbulence-induced curvature term in addition to other aberrations, it is necessary to determine the aberrations induced by such (curved DM surface) an optical element when placed at an angle (other than $0\deg$) of incidence in the optical path. To this effect, we estimate to a first order, the aberrations introduced by a DM as a function of the incidence angle and deformation of the DM surface. We record images using a simple setup in which the incident beam is reflected by a 37 channel Micro-machined Membrane Deformable Mirror for various angles of incidence. It is observed that astigmatism is a dominant aberration which was determined by measuring the difference between the tangential and sagital focal planes. We justify our results on the basis of theoretical simulations and discuss the feasibility of using such a system for adaptive optics considering a trade-off between wavefront correction and astigmatism due to deformation.
\end{abstract}


\maketitle %% required
\section{Introduction}
\label{intro}
It is known that the most important component of an Adaptive Optics (AO) System is the corrector, which is the Deformable Mirror (DM). It can be a continuous face sheet or a surface formed by mirror segments \cite{Roggeman}. In the former the mirror boundary is fixed and voltage applied to any one actuator influences the neighbouring surface as well. In the case of the Micro-machined membrane Deformable Mirror(MMDM) [2] this influence can be as large as 60$\%$ [3]. However, they are still preferred in many AO systems [4-8] because of their low cost and capability of achieving large stroke. A 37 channel MMDM from OKOTECH, Netherlands, is being used by us at the Udaipur Solar Observatory (USO) for its solar adaptive optics [9-11] (SAO) system. Although the MMDM is not common in many SAO systems, its performance has been validated for solar observations[12].

In an AO system, it is necessary to bias the DM in such a way that the stroke can be achieved in both positive and negative directions from the biased position. In general, this can be achieved by applying a constant voltage to all the actuators. In case of the MMDM, application of a constant voltage to all the actuators deforms the mirror to a shape  often approximated to be parabolic [13]. However, such approximations do not hold, at points away from the centre. A DM, whose surface is distorted by applying a uniform set of voltage to its actuators, and placed in the optical setup at an angle to the beam will undoubtedly mimic a tilted parabolic mirror. Such a tilted curved/parabolic mirror will induce optical aberrations such as defocus, astigmatism and coma \cite{Malacara}. While defocus can be compensated by moving the imaging system, astigmatism and coma cannot be corrected by a simple alignment. As the DM is also being used to compensate a turbulence-induced curvature term, it is necessary to determine the maximum angle of incidence at which the DM can be placed in the optical setup without introducing an additional set of aberrations than that which are inherent to the DM. Otherwise, at relatively large angles of incidence, when the DM corrects an atmosphere-induced defocus (curvature term), it will inevitably introduce astigmatism. In order to minimize this astigmatism, an additional correction is required. This constraints the bandwidth of the system.

Along with `setup' induced aberrations, intrinsic aberrations (aberrations due to mirror's surface profile) are also very important. As the technical passport of the DM states that the initial figure of the DM is astigmatic upto 1.3 fringes(P-V)[14], we would like to study the intrinsic aberrations as well.

In this paper we estimate, to a first order, the aberrations introduced in the optical system as a result of folding the beam using a DM with a bias voltage, for various angles of incidence. The rest of the paper is organized as follows. Section 2 describes theoretical aspects related to the deformation of the mirror under an external, uniform voltage and in section 3 we present the theoretical investigation on astigmatism and defocus in such a deformed system. Section \ref{simu} describes the simulations which demonstrate the aberrations introduced when the DM is placed at different angles of incidence. The degradation in the observed image quality with varying curvature for different angles of incidence is discussed in Section \ref{expt} using an experimental setup. The understandings from the simulations and the observations from the experiment are discussed in Section \ref{discu}.
\clearpage
\section{Theoretical perspective of the deformable mirror under the influence of voltage}
\label{dmsag}
\begin{figure}[htbp]
\includegraphics[width=0.2\columnwidth, angle=90]{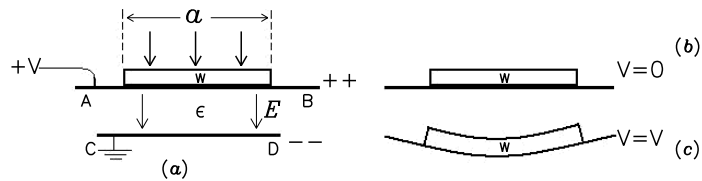}
\caption{Schematic of the membrane surface. W is the wafer surface, $a$ is the diameter of the surface and $\epsilon$ is di-electric constant. $d$ is the distance between the electrodes AB and CD. Thickness of the wafer is $h$.}
\end{figure}

In Figure~1, the silicon wafer W, with upper surface polished is place on the +ve electrode AB, which is subjected to potential $V$, while the -ve electrode CD is grounded. When $V=0$, the wafer W has its undistorted shape $(b)$, while it gets distorted to the shape, shown in $(c)$, when $V \ne 0$. If $d$ be the separation between the plates and $\epsilon$ be the dielectric permittivity of the material, filling the space between the plates, a charge density $\Sigma = \epsilon V/d$ develops on the plates. The resulting stress is given by
\begin{equation}
P = \epsilon V^2/(2 d^2)
\end{equation}
Thus, if $E$ be the young's modulus, $\sigma$ be the Poisson ratio and $h$ be the thickness of the wafer, then the deformation of the upper surface of the wafer is given by\cite{elasti}
\begin{eqnarray}
\label{sageqn}
z(x,y)&=&\frac{-3\,P(1-{\sigma}^2)}{128\,Eh^3}\left[a^2-(x^2+y^2)\right]^2	\\
\textrm{for}\;x=y=0,\quad
z(0,0) &=& -\frac{P}{64}\frac{3}{2}\frac{(1-{\sigma}^2)}{Eh^3}a^4 = -h_0
\end{eqnarray}
where $h_0$ is defined as the sag of the membrane.\\
The gradient of the surface at any point on the $(x , y )$ membrane are given by
\begin{eqnarray}
z_x &=&\partial z/\partial x = \frac{4h_0}{a^4}\:x\left[a^2-(x^2+y^2)\right]	\nonumber	\\
z_y &=&\partial z/\partial y = \frac{4h_0}{a^4}\:y\left[a^2-(x^2+y^2)\right]
\label{zderivate}
\end{eqnarray}
Accordingly, the two curvatures are given for $z_x,\;z_y \ll 1$ 
\begin{eqnarray}
K_x =\frac{1}{\rho_{x}}= -\frac{{\partial}^2 z }{{\partial x}^2} &=& -\frac{4h_0}{a^4}\:\left[a^2-3\:x^2-y^2\right]	\nonumber\\
K_y =\frac{1}{\rho_{y}}= -\frac{{\partial}^2 z }{{\partial y}^2} &=& -\frac{4h_0}{a^4}\:\left[a^2-x^2-3\:y^2\right]
\label{rxry}
\end{eqnarray} 
The convention being that the curvature is positive if the surface is convex (center of curvature is located at $z<0$ i.e below the undistorted surface) and negative if the surface is concave (center of curvature is located at $z>0$ i.e above the undistorted surface). where, $\rho_x$ and $\rho_y$ are the radii of curvature along $x-$ and $y-$ direction respectively. The point where $K_x$ or $K_y$ or both equal to zero indicates the point of inflexion.
Thus, at the center $x=y=0, \rho_x(0,0)=-{a^2}/{4h_0}=\rho_y(0,0)=\rho_0$ and we can express the equation (2) as 
\begin{equation}
z(r) = -h_0+\frac{r^2}{2\rho_0}-\frac{r^4}{4\rho_0a^2}
\end{equation}
where, $r^2 = x^2+y^2$.
In the literature, the formula that is traditionally used for the radius of curvature is $R=({a^2}/{2h_0} + {h_0}/{2})$. This gives a value, which is much larger than the actual radius of curvature at the center, it's magnitude being $\rho_0$. This is because on these formula, the circumference of the mirror $(x^2+y^2=a^2, z)$ and the point of sag $(x= 0 = y, z =-h_0)$ are fitted to lie on a sphere for which simple geometry gives a radius $R$ as given above. The actual values are, however given by equation \ref{rxry}, showing that the surface cannot be described by a unique radius of curvature. The various aberrations, which appear in the Fraunhofer limit are thus to be worked out with $z(x, y)$ being given by equation \ref{sageqn}, in which $z(x,y)$ is found to be proportional to $V^2$, a fact, which we present in our further results. 
\begin{figure}
\includegraphics[width=0.975\columnwidth]{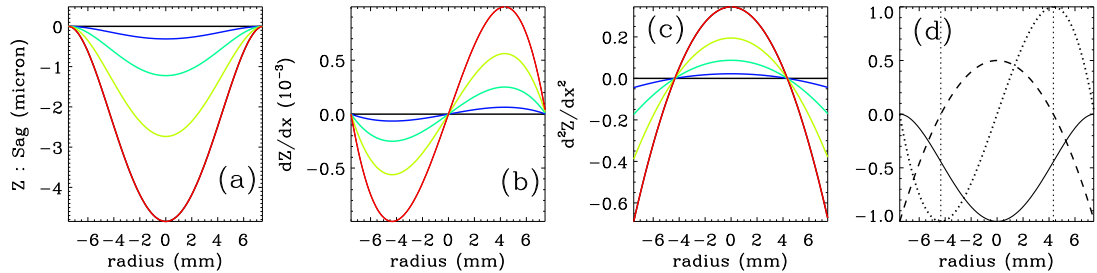}
\caption{(a) Surface elevation $z(x,0)$ (b) gradient $z_x(x,0)$ and (c) curvature of the membrane at different points along the diameter for application of different voltages from 0 to 200~V;(d) elevation $z(x,0)$, (continuous line), gradient (dotted line) and curvature (dashed line), for the application of 200 volts; vertical dotted lines show the points of inflexion. Different parameters for a typical wafer that are being used here are: $E$ = 165 $GPa$, $\sigma$ = 0.24, h=2.5~$\mu$m, $a$=7.5 $mm$. The colors in plots (a-c), black, blue, green, yellow and red represent for voltages 0, 50, 100, 150, and 200~V, respectively.}
\label{sagplots}
\end{figure}

\begin{figure}[!h]
\includegraphics[width=0.4\columnwidth, angle = 90]{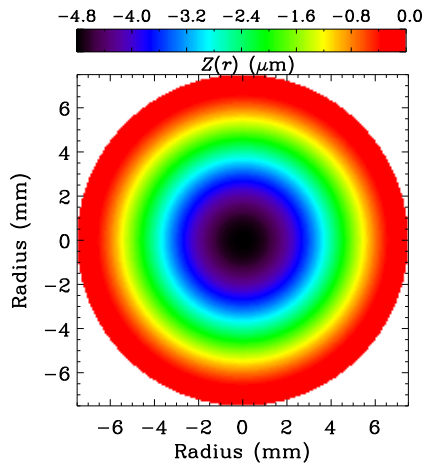}
\caption{ Surface map of the membrane with the application of a voltage of 200~V. We fit this surface with first 21 Zernike polynomials; Z4(defocus) and Z11(Spherical aberration) are the dominant terms.}
\label{sagimg}
\end{figure}

Figure $\ref{sagplots}$ show typical manifestations of $z$, inclination and curvature of the membrane and the location of inflexion for typical values of various parameters of equation \ref{sageqn}. It is clear that the variations in $Z_x$ and $Z_y$ at different points on the surface, imply a reflecting surface with varying inclination, which would, in geometrical viewpoint, reflect the light from these points of incidence in different directions, consistent with the laws of reflection. The curvatures $K_x$ and $K_y$ ascribe focusing and defocusing properties of the reflecting membrane surface. Figure \ref{sagimg} show the surface map of the membrane with the application of voltage; we fitted this surface with 21 Zernike polynomials [16], the retrieved coefficients show that the $Z4$ (defocus) and $Z11$ (spherical aberration) are the dominant terms. 

\section{Image formation by a deformable mirror}
\begin{figure}[t]
\hspace{-1 cm}
\includegraphics[width=0.6\textwidth, angle=90]{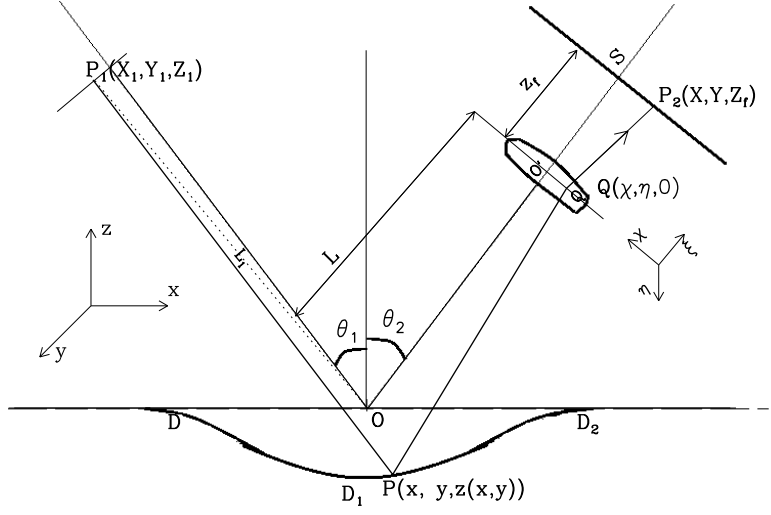}
\caption{Ray diagram for image formation due to deformable mirror}
\label{raydia}
\end{figure}

Figure \ref{raydia} illustrates the image formation by a DM. We assume that the DM lies in the $x\:y$ plane in its un-deformed state, and the curved profile $DD_1D_2$ represents the DM under the influence of uniform voltage. The deformation $z(x,y)$ follows the equation (6) given in section 2. Let a ray from a point source $P_1$(-\x, \y, \z) be incident on the curved mirror at $P(x, y, z(x,y))$. The light ray reflected by the DM is imaged using a lens of focal length ${f_0}$ on to a screen `S' placed at a distance of $Z_f$ from the lens. The lens lies in the $\chi\,\eta$ plane such that $\chi\,\eta\,\zeta$ form a right handed system of orthogonal axes. The lens introduces a path length `Lens' to the ray on propagating through the lens.

In order to find the path followed by the rays, we calculate the path length $P_1\,P\,Q\,P_2$ and minimize the same with respect to $(x, y)$ and $(\chi, \eta)$ as demanded by Fermat's principle. 
The total optical path is given by  
\begin{eqnarray}
L_{tot} &=& {P_1}P+PQ+Q{P_2}+\textrm{Lens}	
\end{eqnarray}
For a perfectly aligned optical system, we have $\theta_1=\theta_2$, so that
\begin{eqnarray}
L_{tot}=\frac{\alpha_{11}}{2}x^2+\frac{\alpha_{22}}{2}y^2+\frac{\beta_{11}}{2}(\chi^2+\eta^2)+\frac{\beta_{22}}{2}(X^2+Y^2)+\gamma_{11}\:x\:\chi+\gamma_{22}\:y\:\eta+\delta_o(X\chi+Y\eta)\nonumber\\
\end{eqnarray}
\begin{eqnarray}
\textrm{where, }\alpha_{11}&=&\cos\:\theta_1\left[\cos\:\theta_1/L-2/{\rho_o}\right]		\nonumber\\
		\alpha_{22}&=&{1}/{L}-{2\:\cos\;\theta_1}/{\rho_o}\nonumber\\
		\beta_{11}&=&{1}/{L}+{1}/{Z_f}-{1}/{f_0}\nonumber\\
		\beta_{22}&=&{1}/{Z_f}\nonumber\\
		\gamma_{11}&=&{-\cos\:\theta_1}/{L},\nonumber\\
		\gamma_{22}&=&{-1}/{L}, \quad\delta_{0}={-1}/{Z_f}
\end{eqnarray}
Here, we have neglected the terms contributing to a constant phase and kept only the leading terms that contribute to astigmatism and defocus. In order to find the path of the ray we apply Fermat's principle. Thus keeping (X, Y) fixed we minimize $L_{tot}$ w.r.t $x, y, \chi, \eta$ respectively, to yield, 
\begin{eqnarray}
\alpha_{11}\:x+\gamma_{11}\:\chi = 0	\nonumber\\
\alpha_{22}\:y+\gamma_{22}\:\eta = 0	\nonumber\\
\gamma_{11}\:x+\beta_{11}\:\chi+\delta_{o}\:X = 0\nonumber\\
\gamma_{22}\:y+\beta_{11}\:\eta+\delta_{o}\:Y = 0
\label{lineqn}
\end{eqnarray}
which are linear equations in the variables x, y, $\chi$, $\eta$, X, Y. Solving the above linear equations,
\begin{eqnarray}
\chi&=&-(\alpha_{11}/\gamma_{11})\:x, \nonumber\\
\eta&=&-(\alpha_{22}/\gamma_{22})\:y	\nonumber\\
X&=&\left[(\alpha_{11}\beta_{11}-{\gamma_{11}}^2)/(\gamma_{11}\delta_o)\right]\,x = \alpha\,x\nonumber\\
Y&=&\left[(\alpha_{22}\beta_{11}-{\gamma_{22}}^2)/(\gamma_{22}\delta_o)\right]\,y = \beta\,y
\label{lineqnsol}
\end{eqnarray}
here, the definitions for $\alpha$, $\beta$ can be found from the pre-factors.
These equations relate ($x, \chi, X; y, \eta, Y$) with each other. In other words, if the diameter of the lens be much larger than $2a$, i.e. the diameter of the DM then for any point of incidence ($x, y, z(x, y)$) on the DM, we know the point ($X,Y,Z_f$), which the ray of light reaches. 

On defining $\tan\,\theta = (x/y)$ we can write the coordinates of the point ($x, y$) on the DM to be $x=r\,\cos\,\theta$ and $y=r\,\sin\,\theta$. Thus, various rays, incident on the circle: $x^2+y^2=r^2$ on the DM, will reach the point ($X(r), Y(r), Z_f$), where $X(r) = \alpha\:r\;\cos\,\theta$, and $Y(r) = \beta\:r\;\sin\,\theta$. This means that $(X(r), Y(r))$ describes an ellipse $(X(r)/\alpha\;r)^2+ (Y(r)/\beta\;r)^2=1$. Thus on making $r=a$ we find all rays will reach $(X_m, Y_m, Z_f)$ where on the screen all the rays will lie inside a ellipse $(X_m, Y_m)$ such that
\begin{eqnarray}
X_m = \alpha\:a\:\cos\,\theta,\quad Y_m = \beta\:a\:\sin\,\theta\\
(X_m/\alpha\:a)^2+(Y_m/\beta\:a)^2 = 1
\end{eqnarray}
The results given in (12) and (13) enable us to estimate the astigmatic aberration inside the system.
\subsection{Location of the Astigmatic foci}
Consider a case where on adjusting the screen and making $Z_f=f_0+\delta_1$ we can make $\alpha=0$. In this case, $X_m=0$ the illumination on the screen will be confined only along the $Y$ axis. The necessary condition for that, i.e $\alpha=0$ requires $\alpha_{11}\,\beta_{11}={\gamma_{11}}^2$, as seen from equations (11). Then on using the definition given in equation (9), we find in the limits $\left|\delta_1/f_o\right|<<1$ and $\rho_0<<2\;L$, 
\begin{eqnarray}
\delta_1&=&-(2/\rho_0\cos\,\theta_1){f_0}^2, \nonumber\\ 
Z_f&=&f_1=f_0+\delta_1, \nonumber\\ 
&=&f_0-(2{f_0}^2/\rho_0\cos\,\theta_1)
\end{eqnarray}
Similarly on adjusting $Z_f=f_0+\delta_2$ we can make $\beta=0$. In this case $Y_m=0$ and the illumination on the screen is confined only along the axis. For this to happen, we must have $\alpha_{22} \beta_{11}={\gamma_{22}}^2$ so that for $\left|\delta_2/f_o\right|<<1$, we get on using the definitions in equations (9),
\begin{eqnarray}
\delta_2&=&-(2\,\cos\,\theta_1/\rho_0){f_0}^2, \nonumber\\
Z_f&=&f_2=f_0+\delta_2, \nonumber\\
&=&f_0-(2\,\cos\,\theta_1\:{f_0}^2/\rho_0)
\end{eqnarray}
The astigmatism in the system is then estimated as 
\begin{eqnarray}
\delta\,f&=&f_1-f_2=-2\,{f_0}^2\,\sin^2\,\theta_1/\rho_0\,\cos\,\theta_1 \nonumber\\
&=&\frac{-2{f_0}^2}{\rho_0}f(\theta_1) \propto\;V^2{\theta^2}_1
\end{eqnarray}

The above equation for astigmatism ($\delta f$) shows a strong dependence on $\theta_1$, it varies very fast for $\theta_1\ge 60^\circ$ and blows up as $\theta_1\rightarrow90^\circ$ (c.f. Table 1). It also shows that $\delta f$ varies as $V^2$.

\subsection{Location of circle of least confusion}
For any value of $Z_f=f_0+\Delta$ the patch of light on the screen lies inside an ellipse, defined by equations (12) and (13). Thus, the extent of the patch is defined by averaging $R_m^2(\theta) = X_m^2(\theta)+Y_m^2(\theta)$ over $0\le\theta\le 2\pi$. These give, 
\begin{eqnarray}
\overline{R_m^2(\theta)} &=& \overline{X_m^2(\theta)}+\overline{Y_m^2(\theta)}\nonumber\\
&=& \alpha^2 a^2\,\overline{\cos^2\theta} + \beta^2 a^2\,\overline{\sin^2\theta}\nonumber\\
&=& \frac{a^2}{2}\left[\alpha^2+\beta^2\right]
\end{eqnarray}

The circle of least confusion is located at a point, where $\overline{R_m^2}$ is minimum, i.e. at a value of $\Delta$ for which, 
\begin{eqnarray}
\frac{d}{d\Delta}\overline{R_m^2} = 0 =\frac{a^2}{2}\left[2\alpha \frac{d\alpha}{d\Delta}+2\beta\frac{d\beta}{d\Delta}\right] 
\end{eqnarray}
\\
These derivatives can be evaluated from equations (11) on using the definitions given in equation (9). We find that in the limit $\rho_0\cos\theta_1\rightarrow\infty$, i.e. for curvatures not too large, as is the case for most practical situations. 
\begin{eqnarray}
\Delta\approx-(4f_0^2/\rho_0)\,F(\theta_1)\propto\,V^2\,F(\theta_1)\\
\textrm{where, }F(\theta_1)= \frac{\cos\theta_1}{1+\cos^2\theta_1}
\end{eqnarray}

\begin{table}
\caption{Dependence of circle of least confusion on angle of incidence}
\begin{center}
\begin{tabular}{|c|c|c|c|c|c|c|c|}
\hline
$\theta_1$&0&15&30&45&60&75&90\\
\hline
$f(\theta_1)$&0.000&0.069&0.288&0.707&1.500&3.605&$\infty$\\
$F(\theta_1)$&0.500&0.499&0.495&0.471&0.400&0.242&0\\
\hline
\end{tabular}
\end{center}
\end{table}

We find that $F(\theta_1)$ varies very slowly with $\theta_1$ until $\theta_1$ exceeds $60^\circ$. Beyond this the condition of large $\rho_0\cos\,\theta_1$ may breakdown and the above approximation is no longer valid. Variation of $F(\theta_1)$ with respect to $\theta_1$ is shown in Table~1. However for low values of $\theta_1$ the point of least confusion shifts as given by equation (20), being quadratic in V and weakly dependent on $\theta_1$. The results given above are tested by simulation and experimentally and are described in the subsequent sections. 

It is to be noted that in our analysis we have kept in Eq.(8) terms which are up to the second order in x, y, $\chi$, $\eta$, X, and Y.  Extending to a higher order, there appears a term $2(\chi X+\eta Y)(X^2 + Y^2)/Z_f^3$ in $L_{tot}$ that is ascribed to aberration due to coma, which we neglected in our analysis for the following reason. 

The extra term gives rise to additional terms linear in X and Y in the last two equations of  Eq.(10). These additional terms are smaller than the last terms in these equations ( $\delta_0$X and $\delta_0$Y respectively, with $\delta_0 = 1/Z_f$) by a factor, which is of the order of $(a/Z_f)^2 \approx$ 1.4$\times 10^{-3}$ as can be seen with a= 7.5 mm and $Z_f \approx f_0$= 200 mm as is the case in our system. This smallness of coma in comparison to astigmatism is also borne out by the results displayed in Table 3.

\section{Simulations}
\label{simu}
The simulations were carried out using $ZEMAX{^{TM}}$ [15], an optical design software to study the influence of voltage on the DM kept at an angle to the incident beam. The simulations were performed for the DM with and without any intrinsic surface figure. 

\begin{figure}[!b]
\includegraphics[width=7cm]{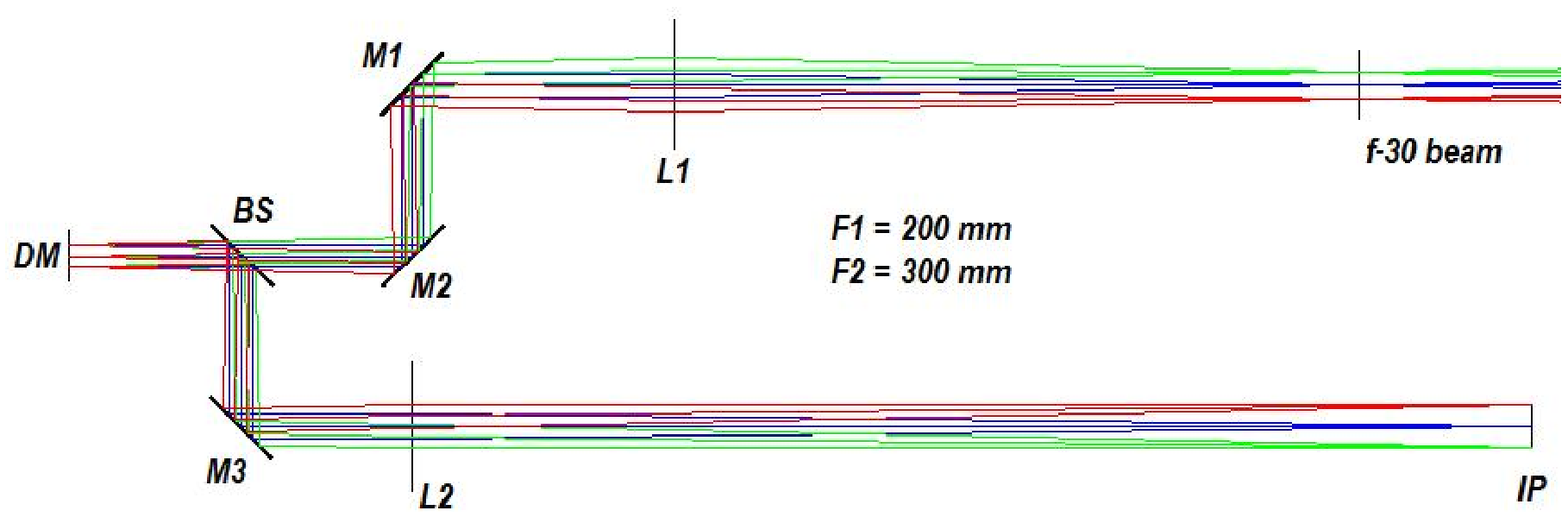}
\includegraphics[width=7cm]{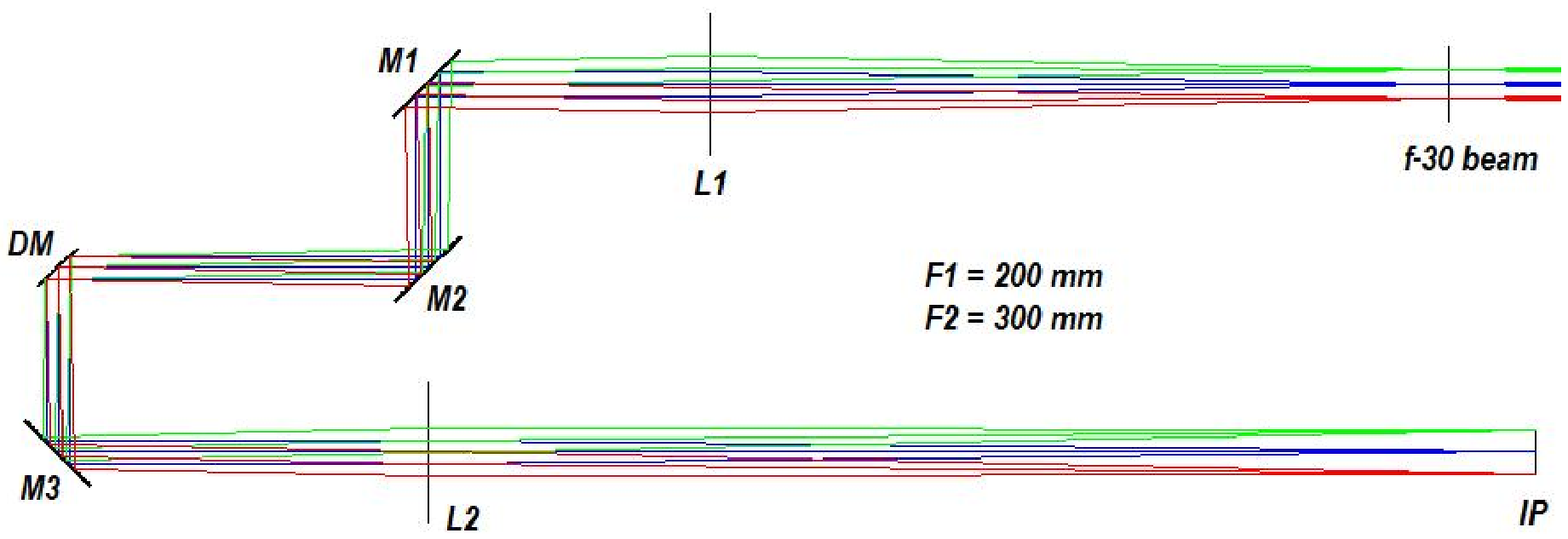}
\caption{Optical layouts of the two configurations; Left: configuration-1, Right: configuration-2. An Input beam of F\#30 was collimated by a lens of focal length 200 mm. M1 and M2 are steering mirrors, one of them was used as Tip-tilt mirror (TTM). DM was placed at the pupil plane. The collimated light was focused using a lens (L2) of focal length of 300 mm, which gives a plate scale of 30.5 arc-sec/mm at the image plane (IP). In configuration-1, one of the reflected beams from beam splitter (BS) is not shown here.}
\label{od}
\end{figure}

\subsection{Simulations without any intrinsic aberrations}
\label{simIntri}
Figure \ref{od} describes the two optical configurations (OC) where the DM is placed at 0$^\circ$ (OC-1) and 45$^\circ$ (OC-2) to the incident collimated beam, respectively. The radius of curvature ($\rho_0$) of the DM is related to its radius $(a)$ and sag $(h_0)$ by the relation
\begin{equation}
\rho_0 = \frac{a^2}{4h_0} = \frac{256\:\:h^3\:d^2}{3\epsilon (1-{\sigma}^2)\:a^2\: V^2}\propto \frac{1}{a^2V^2}
\end{equation}

The radius of curvature ($\rho_0$) was calculated for voltages from 0 to 225 volts in steps of 25 volts. It is to be noted here that the user manual supplied by the manufacturer of the DM states that the maximum central deflection of the DM surface is 7~${\mu}m$. As a result, the f-number of the DM varies from infinity to $\approx$ 67.
 %Figure 2(a) shows the change in the radius of curvature and the sag of the DM with the application of voltage. 
The final wavefront quality of the system was estimated by assigning the values of $\rho_0$ to the DM in OC-1 and OC-2. In OC-1, application of voltage to DM (i.e. RC to DM) causes only defocus error, whereas in case of OC-2, it also causes astigmatism in addition to defocus. We estimated the aberration coefficients in terms of Zernike coefficients.

\begin{table*}[t]
\begin{center}
\caption{Aberration coefficients of Configuration-OC1. The Zernike coefficients before and after correspond to the cases when a voltage was applied to the DM that was then followed by shifting the focal plane during the aberration minimization scheme.}
\label{info1}
\begin{tabular}{cccccccccc} 
\hline
Voltage& Sag & $\rho_0 $ &  Defocus  & \multicolumn{4}{c}{Zernike Coefficients}&\multicolumn{2}{c}{WFE ($\lambda$)}\\
\cline{5-8}
\cline{9-10}
V&({$\mu$}m)&$(m)$& $(mm)$ & \multicolumn{2}{c}{Z4} & \multicolumn{2}{c}{Z11$\times10^{-5}$}&Before&After\\
\cline{5-8}
\cline{7-8}
&&&& Before & After & Before & After&&\\
&&&&&$\times10^{-5}$&&&&$\times10^{-4}$\\
\hline
0   & 0.00  &  $\infty$   & 0      & 0    & 0   & 0    & 0 & 0    & 0 \\
25  & 0.07  &  187.6 & 001.71 & 0.07 & 0.4 &0.21  & 0 & 0.07 & 0.06\\
50  & 0.30  &  46.90 & 006.85 & 0.28 & 1.6 &0.84  & 0 & 0.28 & 0.24\\
75  & 0.67  &  20.84 & 015.50 & 0.63 & 3.6 &1.87  & 0 & 0.63 & 0.55\\
100 & 1.20  &  11.72 & 027.77 & 1.12 & 6.4 &3.28  & 0 & 1.12 & 0.97\\
125 & 1.87  &  07.50 & 043.82 & 1.75 & 10.0&5.02  & 0 & 1.75 & 1.52\\
150 & 2.70  &  05.21 & 063.86 & 2.52 & 14.4&7.01  & 0 & 2.52 & 2.18\\
175 & 3.67  &  03.83 & 088.20 & 3.43 & 19.6&9.17  & 0 & 3.43 & 2.97\\
200 & 4.80  &  02.93 & 117.18 & 4.47 & 25.6&11.35 & 0 & 4.47 & 3.89\\
225 & 6.07  &  02.31 & 151.25 & 5.66 & 32.3&13.35 & 0 & 5.66 & 4.93\\
\hline
\end{tabular}
\end{center}
\end{table*}

\begin{table*}
\begin{center}
\caption{Aberration coefficients of Configuration-OC2}
\label{info2}
\begin{tabular}{cccccccccccc} \hline
Voltage & Defocus & \multicolumn{6}{c}{Zernike Coefficients}&\multicolumn{2}{c}{WFE ($\lambda$)}\\
\cline{3-7}
\cline{8-10}
(V)&$(mm)$ & \multicolumn{2}{c}{Z4} & \multicolumn{2}{c}{Z6}&\multicolumn{2}{c}{Z7$\times10^{-4}$}&Before&After\\
\cline{3-8}
&&Before & After & Before & After& Before&After&&\\
&&&$\times10^{-4}$&&&&&&\\
\hline
0   & 0     &   0  & 0    & 0     & 0     & 0    & 0     & 0    & 0\\
25 & 01.18 & 0.07 & 0.11 & 0.035 & 0.035 & 0.01 & 0.008 & 0.08 & 0.04 \\
50 & 07.27 & 0.29 & 0.44 & 0.139 & 0.139 & 0.18 & 0.137 & 0.33 & 0.14 \\
75 & 16.45 & 0.67 & 0.99 & 0.315 & 0.315 & 0.92 & 0.694 & 0.74 & 0.32 \\
100 & 29.49 & 1.19 & 1.77 & 0.559 & 0.559 & 2.92 & 2.193 & 1.31 & 0.56 \\
125 & 46.56 & 1.85 & 2.76 & 0.874 & 0.874 & 7.13 & 5.353 & 2.05 & 0.87 \\
150 & 67.91 & 2.67 & 3.97 & 1.259 & 1.259 & 14.8 & 11.10 & 2.95 & 1.26 \\
175 & 93.88 & 3.63 & 5.39 & 1.714 & 1.714 & 27.4 & 20.57 & 4.02 & 1.71 \\
200 & 124.8 & 4.75 & 6.99 & 2.238 & 2.238 & 46.7 & 35.08 & 5.25 & 2.24 \\
225 & 161.4 & 6.00 & 8.76 & 2.833 & 2.833 & 74.9 & 56.20 & 6.65 & 2.83 \\
\hline
\end{tabular}
\end{center}
\end{table*}

In OC-1, the dominant aberration is defocus $(Z4)$ along with a negligible amount of spherical aberration$(Z11)$ of the 3rd order (Table \ref{info1}). These aberrations increase with the increase in voltage as seen from their corresponding Zernike coefficients (cf., Table \ref{info1}). To compensate these aberrations the focal plane position must be shifted which is achieved by minimizing the merit function for an optimal focal plane position in ZEMAX. The process of optimization yields a wavefront error which is well within the diffraction limit.

In OC-2, the dominant aberrations were defocus (Z4), astigmatism (Z6) along with small amounts of coma (Z7) (Table \ref{info2}). These aberrations also increase with an increase in voltage. After optimizing the focal plane position, the Z4 coefficient reduces considerably, while the coefficient Z6, namely astigmatism, remains unchanged. The overall wavefront quality varies from $\lambda$/12.5 to 6.6$\lambda$ when the voltage changes from 25-225 $V$, before optimization. After optimization the values vary from $\lambda$/25 to 2.83$\lambda$ for the same range of voltage. Thus at an angle of incidence of 45$^\circ$, application of any voltage to the DM introduces an additional astigmatism and coma (cf., Table 3). It is to be noted that here the coefficient of coma is very small in comparison to that of astigmatism. Hence, we neglect coma in our further analysis. However, we conjecture that the coefficient of coma is negligible because of the large f-number of the curved DM mirror, this may not be the case with a tilted curved mirror with a smaller f-number, where the coma is equally important as astigmatism \cite{Malacara}.

\begin{figure}[ht]
\centerline{\includegraphics[width=8cm,angle=0]{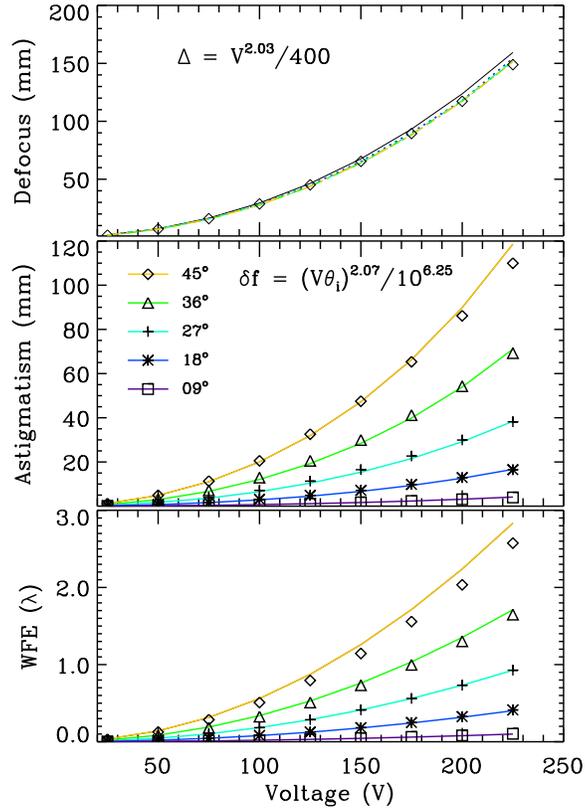}}
\caption{Top panels show amount of focus shift needed to minimize the aberrations with the application of voltage at various angles of incidence. Middle panels show the change in the amount of astigmatism with the applied voltage for different angles of incidence ($\ti$). Bottom panels: Wavefront error (WFE) after optimization for optimum focus. Relation between applied voltage and astigmatism and defocus are shown in the respective figures. The symbols represent the fitted values based on relations derived empirically.}
\label{intri}
\end{figure}

As a further step, the amount of astigmatism induced by the (curved) DM for different angles of incidence was studied, as it is the only term that remains unaltered after optimization. As a matter of convenience the astigmatism is expressed as the change in tangential and sagital focal planes rather than Zernike coefficients as these values can be compared with those from an optical setup having a test target. Figure \ref{intri} shows the increase in astigmatism (change in sagital and tangential focii) with increase in voltage (i.e decrease in RC) for different angles of incidence. Dependence of astigmatism and defocus on voltage and angle of incidence is derived empirically. It shows that defocus depends only on the voltage applied, whereas astigmatism depends both on the voltage applied and the angle of incidence. This is in agreement with the theory (cf., equations 16 and 19). Similarly, the wavefront error of the system at optimum focus varies with angle of incidence and it shows a similar trend as astigmatism. The system shows a considerable amount of WFE for angles of incidence $\ti\geq$ 9$^o$, with the WFE is worse than $\lambda$/10 for any voltage.  

\subsection{Simulations with intrinsic astigmatism}
The technical document of the DM states that in the absence of any voltage the initial mirror figure is astigmatic upto 1.3 fringes. In the simulations, this intrinsic aberration was modeled by utilizing ``Zernike Fringe Sag"  [17]. The terms Z5, Z6 and, Z12, Z13 refer to the first and third order astigmatism, respectively and were adjusted such that the initial difference between the tangential and sagital focus was 6 mm which corresponds to a wavefront error of 0.18~$\lambda$ at $\lambda$ = 550 nm. The obtained coefficients were assigned to the DM to make the surface astigmatic. The entire exercise as described in Section~\ref{simIntri} was repeated, astigmatism and wavefront error were estimated for different angles of incidence. At an angle of incidence of 0$^\circ$, the wavefront error remains constant at 0.18~$\lambda$ after optimizing the focal plane for different values of RC, which correspond to the intrinsic astigmatism. At any other angle of incidence the wavefront error shows quadratic variation with the increase in RC with a minimum value being around 0.18~$\lambda$. This shows that the astigmatism arising from the surface figure needs to be corrected before using the DM in an optical setup. 

\section{Experiment}
\label{expt}
\subsection{Optical Setup}
To compare the results from the ZEMAX simulations, we carried out an experiment with the DM, using an F\#15 Coude telescope as the light feed. With a set of relay lenses the image was magnified by a factor of 2. At the focal plane of the F\#30 beam an artificial target was placed which was illuminated by sunlight. A lens of focal length 200 mm was used to collimate the light modulated by the target. The DM was placed in the collimated beam which reflects the beam towards an imaging lens of focal length 300 mm (refer figure \ref{od}). The voltage to the DM was controlled by 2, 8-bit PCI cards whose maximum output voltage to any channel is 5 V. The individual actuators were first assigned a port address as stated in the user manual and the PCI output was checked for each channel. A high voltage amplifier consisting of 2 high voltage amplifier boards, boost the signal from the PCI cards. Each amplifier board contains 20 non-inverting DC amplifiers with a gain of 59. A high voltage stabilized DC supply was used to power the amplifier boards. The maximum operating voltage of the DM is 162 V which corresponds to 256 Digital-to-Analog Counts (DAC). A $1392\times1024$ pixel, Cool-Snap HQ CCD with a pixel size of 6.45 $\mu$m from Roper Scientific was used to image the target. The difference between tangential and sagital planes and the circle of least confusion (optimum focus) were measured in order to estimate the intrinsic as well as induced aberrations.  
\begin{figure}[t]
\centerline{\includegraphics[width=0.7\textwidth]{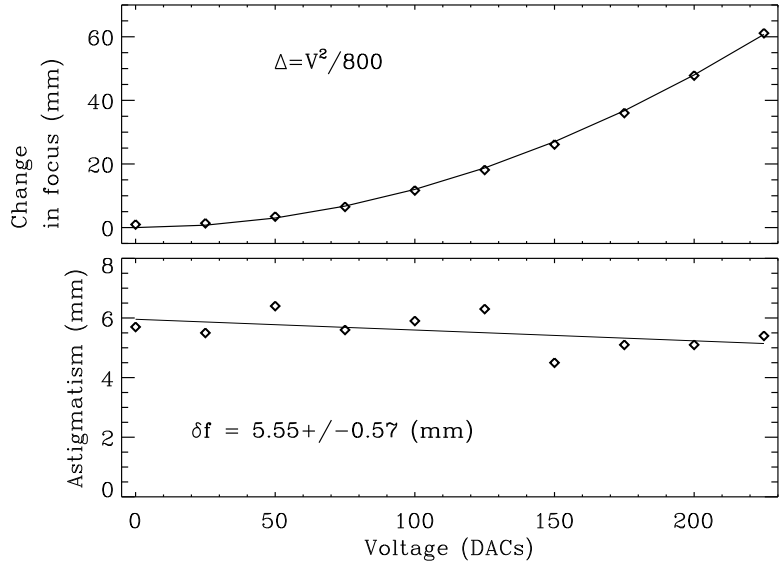}}
\caption{Defocus and astigmatism when the DM is kept at angle of incidence of 0$^\circ$. Top panel show change in focus w.r.t change in voltage. Bottom panel show Intrinsic astigmatism, as there is no tilt in the DM, astigmatism observed is intrinsic to the DM. Solid line represents the linear fit corresponds to the data points.}
\label{expt0}
\end{figure}

\subsection{Intrinsic aberrations and Image quality}
For measuring the intrinsic aberration the DM was placed at an angle of 0$^\circ$ similar to OC-1 in Figure \ref{od}. The target had an L shape pattern and it was observed that the vertical line was focused at one location while the horizontal line at another. This difference in the sagital and tangential focus is caused by astigmatism. To see if this astigmatism changes with the curvature of the DM, a uniform voltage was applied to all 37 actuators of DM from 0 to 225 DACs in steps of 25 DACs. For every voltage set, the tangential, sagital and optimum focal positions ($f_1$ and $f_2$) were measured and the corresponding images were recorded. The difference between the sagital and tangential focal positions do not vary with voltage (cf., Figure \ref{expt0}) which is in agreement with the results from the theory and simulation when an intrinsic astigmatism is considered for 0$^\circ$ angle of incidence. The optimum focus changes quadratically with the voltage applied.
\begin{figure}[t]
\centerline{\includegraphics[width=0.2\textwidth,angle=90]{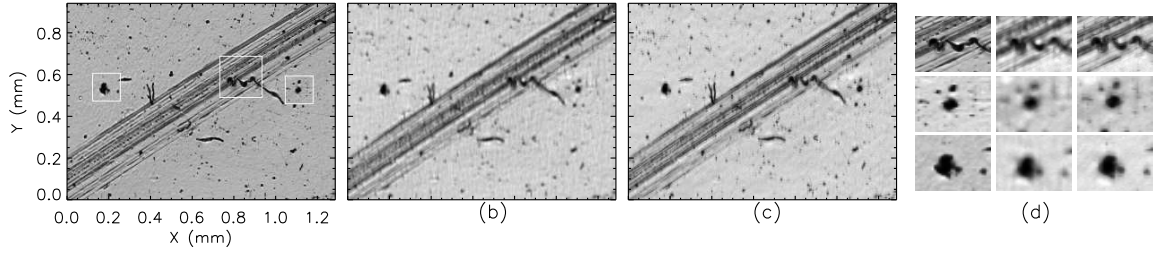}}
\caption{Image quality (a) due to plane mirror (along with three features identified for further processing), contrast = 1, (b) due to DM (plane mirror replaced by a DM), contrast = 0.6, (c) after applying some correction to DM, contrast = 0.64.(d) Three rows corresponds to three features identified in (a) shown in white boxes; along the row the images are due to plane mirror, DM and correction respectively; contrast improvement varies from 4\% to 10\%. The scale shown is in mm.}
\label{images0}
\end{figure}

\begin{figure}[t]
\centerline{\includegraphics[width=0.5\textwidth,angle=90]{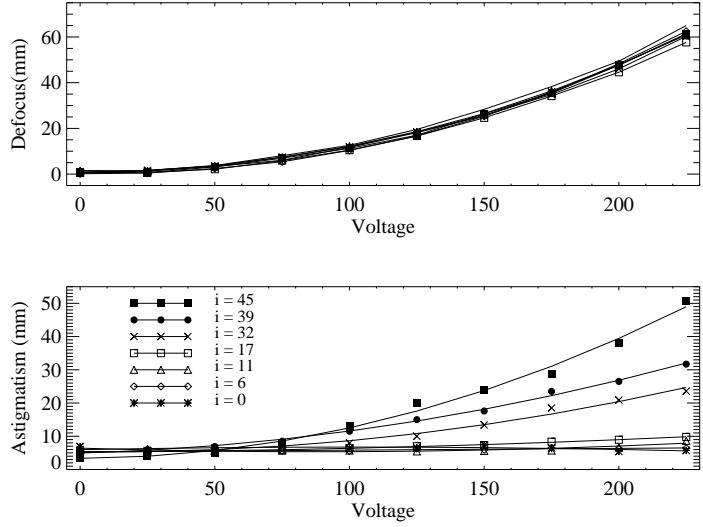}}
\caption{Top panel shows the change in focal plane (circle of least confusion) with the increase in voltage (in DACs). Bottom panel shows astigmatism with increase in in voltage (in DACs) for different angles of incidence.}
\label{expt45}
\end{figure}

\begin{figure}[t]
\centerline{\includegraphics[width=11cm,angle=90]{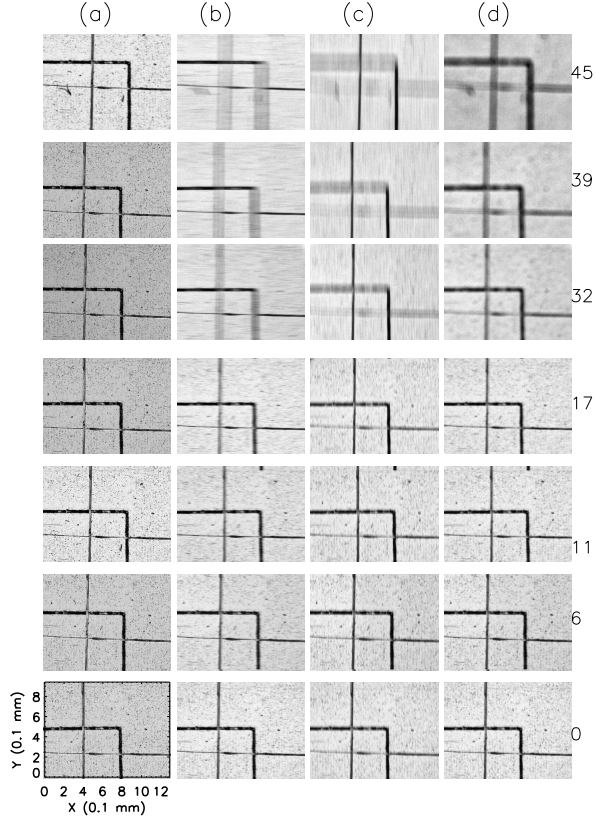}}
\caption{Images at 225 DACs to all actuators of DM for different angles of incidence (a) due to plane mirror, (b) due to DM at sagital plane (c) due to DM at tangential plane (d) due to DM at the plane of least confusion. The images are of size 300 $\times$ 220 pixels$^2$ with a scale of 4.3 micron per pixel.}
\label{images45}
\end{figure}

The presence of an intrinsic astigmatism results in a poor image quality which can be judged visually as shown in Figure \ref{images0} which also shows images taken by a plane mirror kept at the same location of the DM. Astigmatism essentially arises due to different curvatures along different directions, hence as a first order trial we applied maximum voltage to a specific line of actuators alone and recorded the corresponding image as shown in Figure \ref{images0}c. As a performance measurement, contrast of the images was estimated before and after correction for the entire image as well as for a few selected features as shown in Figure \ref{images0}d. Although the improvement in image contrast is nominal, it is evident that a specific voltage set is necessary to compensate for the degradation caused by the intrinsic astigmatism.

The experiment was repeated by placing the DM at several angles of incidence and the tangential, sagital and optimum focus positions ($f_1$ and $f_2$) were measured and the corresponding images were also recorded. The defocus is quantified by noting the distance $\Delta$ by which the screen has to be shifted to reach the position of the circle of least confusion. The astigmatic aberrations is quantified by $\delta\;f = \left|f_1-f_2\right|$. Both $\Delta$ and  $\delta\;f$ are proportional to $V^2$ , their dependence on $\theta_1$ are strikingly different as seen from equations 16 and 20. While the defocus $\Delta$ has a weak dependence on $\theta_1$, that of $\delta\;f$ has a strong dependence on $\theta_1$ as seen from the functional forms of $F(\theta_1)$ and $f(\theta_1)$.

The results are displayed in Figure \ref{expt45} which is in agreement with those from the theory and simulations. Interestingly, for angles less than 10$^\circ$ the astigmatism remains nearly a constant with the applied voltage as seen from the theory and simulations. Images obtained for different angles of incidence by applying 225 DACs to all the actuators of DM are shown in Figure \ref{images45}.

\section{Discussion and conclusions}
\label{discu}

A deformable mirror is an important component in an adaptive optics system for compensation of atmospheric turbulence. A simple optical setup was made to study alignment issues and the intrinsic aberrations of a deformable mirror without employing any kind of wavefront sensor. 

The nature of the surface deformation under the influence of a uniform voltage to all the actuators of deformable mirror is obtained theoretically and given in equation (2). The equation is used to calculate the sag of the deformed/curved mirror. It is shown that radius of curvature of the deformable mirror is inversely proportional to the square of the voltage applied. Furthermore, we also analytically estimated the defocus and astigmatism due to such a curved mirror, which are shown in equations (16) and (20). 

Simulations were performed with different voltages to the deformable mirror, which is kept at different angles of incidence. It is demonstrated that the estimated error in focus is solely a function of the applied voltage where the former has a quadratic dependence on the latter. The same is seen in astigmatism as well with an additional non-linear dependence on the angle of incidence. Simulations also shows that coefficient of coma is negligible in comparison to astigmatism, could be due to the large f-number of DM.

Similar results were obtained from the experiment, wherein a 37 channel MMDM is placed in the collimated beam at different angles of incidence. The DM when placed at 0$^\circ$ angle of incidence shows a finite amount of intrinsic astigmatism which does not vary significantly with the radius of curvature (i.e. application of voltage to DM); this is in agreement with the technical report provided by the manufacturer. It is also observed that the optimum focal plane position changes quadratically with voltage and does not exhibit a change of sign which concludes that the DM does not possess any intrinsic curvature that would change sign on application of voltages. We have also shown that the image quality degrades with the application of voltage and the astigmatism increases with the increase in angle of incidence. 

From both, simulations and experiment it is demonstrated that when the DM is kept at angles greater than $0{^\circ}$ and voltages are applied to it, there is an induced astigmatism which increases in general with the angle of incidence as is also predicted by the theory. The close correlation between theory and experiment enable us to reach the following conclusion about the performance of DM. It shows that apart from the forces appearing due to charging of the capacitor, there are no other spurious deforming forces in the system and the optics of the DM can thus be regarded to have sufficient reliability. The detailed intensity distribution on the screen can also be computed by using equation (9) for the phase distortion due to deformation.

It  is concluded that in order to operate the DM in real-time to correct for atmospheric induced aberrations, it necessary to a) keep the DM at a very small angle of incidence and b) to derive a suitable voltage set that will compensate the intrinsic astigmatism as well. This voltage set can be simply added to any constant voltage set required to bias the DM surface, since the intrinsic astigmatism is independent of voltage; it naturally facilitates using a suitable bias voltage in order to allow the membrane to move in either direction, which is being pursued and will be reported in future.

In practical situations, for wavefront correction the deformation may not be as simple as considered in the present investigations. However, the present study enables us to ascertain the inherent defects in the system which has to be kept in mind, while using the deformable mirror for phase distortion correction. 

Rohan E. Louis is grateful for the financial assistance from the German Science Foundation (DFG) under grant DE 787/3-1 and the European Commission's FP7 Capacities Programme under Grant Agreement number 312495.

\end{document}